\documentclass[ aps,%
floatfix,%
final,%
notitlepage,%
oneside,%
onecolumn,%
nobibnotes,%
nofootinbib,%
superscriptaddress,%
showpacs,%
]%
{revtex4}

\usepackage{graphicx}
\newcommand{\eqref}[1]{(\ref{#1})}

\begin{document}

\title{Scalar mesons in $\eta'\to\eta\pi^{0}\pi^{0}$ decay}

\author{S.V. Donskov}
\author{A.K. Likhoded}\email{Likhoded@ihep.ru},
\author{A.V. Luchinsky}\email{Alexey.Luchinsky@ihep.ru},
\author{V.D. Samoylenko}

\affiliation{Institute for High Energy Physics, Protvino, Russia}

\begin{abstract}
The decay $\eta'\to\eta\pi^{0}\pi^{0}$ is studied in the framework of isobar model. It is shown, that good agreement with the experiment is achieved if $a_0$- and $\sigma$-meson contributions are taken into account. The contribution of $a_0$-meson is dominant, but $\sigma$-meson is necessary to reproduce the form of experimental Dalitz plot. Instead of  usual Breit-Wigner form of $\sigma$-meson propagator we use parametrization of $\pi\pi$-amplitude, that satisfies analiticity, crossing, unitarity and chirality constraints. This amplitude has a pole in complex plane, that corresponds to $\sigma$-meson and describe experimental data on $\pi\pi$-scattering in $K_{e4}$ decay.
\end{abstract}


\maketitle

\section{Introduction}

Precision measurement of the decay

\begin{equation}
\eta'\rightarrow\eta\pi^{0}\pi^{0}\label{epp}
\end{equation}
starts with the work \cite{matrix}, where parameters of the matrix element of this decay in linear approximation were first determined with high accuracy. In this work squared matrix element $\left|\mathcal{M}\right|^{2}$ is expanded in Dalitz variables
\begin{equation}
X=\frac{\sqrt{3}}{Q}\left(
    T_{\pi_{1}^{0}}-T_{\pi_{2}^{0}}
    \right),\qquad
Y=\left(2+\frac{m_{\eta}}{m_{\pi^{0}}}\right)\frac{T_{\eta}}{Q}-1,\label{XY}
\end{equation}
where $T_{\pi_{1}^{0}},T_{\pi_{2}^{0}},T_{\eta}$ are kinetic energies of $\pi^{0}$- and $\eta$-mesons in $\eta'$-meson rest frame $(T_{\pi_{1}^{0}}>T_{\pi_{2}^{0}})$ and $Q=T_{\pi_{1}^{0}}+T_{\pi_{2}^{0}}+T_{\eta}=m_{\eta'}-m_{\eta}-2m_{\pi^{0}}$. Now parametrization with quadratic terms proposed in \cite{schechter} is widely used:
\begin{eqnarray}
|M|^{2}\propto1+a Y+b Y^{2}+d X^{2},\label{general}\label{eq:M2}
\end{eqnarray}
where $a,b,c,d$ are real numbers.

This approach was used in subsequent works studying the decay (\ref{epp}) in neutral \cite{neutral} and charged \cite{charged} modes. In both cases one has good agreement in coefficients of expansion (\ref{general}) and non-vanishing quadratic terms.

The form of the expansion (\ref{eq:M2}) is motivated by chiral perturbation theory (ChPT). In the framework of this model the process $\eta'\to\eta\pi\pi$ was studied theoretically in numerous works \cite{Cronin:1967jq,Schwinger:1968zz,Majumdar:1968,Di Vecchia:1980sq,Tornquist}. It turns out, however, that leading order ChPT calculations gives small value of partial width of this decay and overestimate the slope parameters. The reason for this discrepancy is that the mass of $\eta'$-meson large, so leading ChPT approximation cannot be used. Final state interaction, on the other hand, leads to significant contribution of scalar meson resonances $a_0$, $\sigma$, $f_0$, etc. In ChPT these contributions are taken into account introducing additional contact terms. The resulting width, however, is smaller than the experimental value. In works \cite{Singh:1975aq} it was shown that this problem can be solved in the framework of isobar model with contributions of $a_0$, $\sigma$ and $f_0$ mesons taken into account. This conclusion was also confirmed in other works \cite{Deshpande:1978iv,shecter1,schechter}.

In our paper we use this approach to analyze new experimental data on $\eta'\to\eta\pi^0\pi^0$ reaction \cite{na48}. It is well known, that $\sigma$-meson contribution can not be described with usual Breit-Wigner parametrization. Analyticity, unitarity and chiral properties of $\pi^0\pi^0$ scattering, on the other hand, put strong constraints on the possible form of the amplitude of this process. In the works \cite{Yndurain:2007qm,Caprini:2008fc} the parametrization of the amplitude, that satisfies all mentioned above properties, was presented. Free parameters of this parametrization are fixed from the values of $S$-wave $\pi\pi$ scattering length and experimental data taken from $K_{e4}$ decay (i.e.  $K\to\ell\bar\nu\pi\pi$). This is the main difference of our approach in comparison with the other works, where simple Breit-Wigner form of $\sigma$-meson propagator was used.
Since we use complete amplitude of $\pi\pi$-scattering in our kinematic region, contributions from other resonances (for example, $f_0$-mesons) are also included automatically.
In addition, for the propagator of $a_0$-meson we use slightly modified expression, that takes into account the closeness of this meson to $\eta\pi$ and $KK$ thresholds.

The rest of the paper is organized as follows. In the next section we give parameterizations for amplitudes of final meson interaction in $\pi\pi$ and $\pi\eta$ channels. In section III the results of the fits of these parameterizations on experimental Dalitz plot of $\eta'\to\eta\pi\pi$ decay are presented. Discussion of our result is given in the conclusion.

\section{Matrix Element}

\begin{figure}
\includegraphics{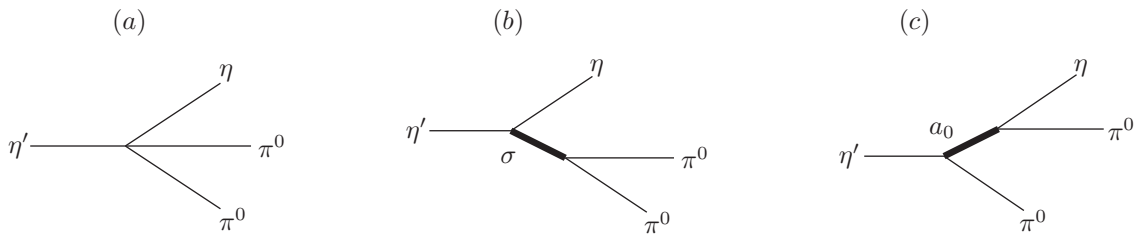}
\caption{Typical diagrams for the process  $\eta'\to\eta\pi\pi$ \label{diags}}
\end{figure}

In the framework of isobar model the matrix element of $\eta'\to\eta\pi^{0}\pi^{0}$ decay can be described by taking into account the contributions of nearest scalar resonances: $a_0$ meson in $\pi\eta$-channel and $\sigma$-meson in $\pi\pi$-channel (see fig.\ref{diags} for typical diagrams). It should be noted, that there are also scalar $f_0$-meson that can give contribution to $\pi\pi$-scattering amplitude (diagram \ref{diags}b). In the present article we use total amplitude for $\pi\pi$-scattering in our kinematic region, extracted from experimental data. It is clear, that there are not only $\sigma$-meson, but also $f_0$-meson contributions in this amplitude.


%

The amplitude of the process $\eta'\to\eta\pi\pi$ can be written in the form
\begin{eqnarray*}
\mathcal{A} & = & \mathcal{A}_{\pi\eta}\left(s_{1}\right)+\mathcal{A}_{\pi\eta}\left(s_{2}\right)+\mathcal{A}_{\pi\pi}\left(s_{3}\right),
\end{eqnarray*}
where $\mathcal{A}_{\pi\eta}$ and $\mathcal{A}_{\pi\pi}$ are rescattering
amplitudes in $\pi\eta$- and $\pi\pi$-channels respectively and
\begin{eqnarray*}
s_{1} & = & \left(p_{2}+p_{3}\right)^{2},\qquad s_{2}=\left(p_{1}+p_{3}\right)^{2},\qquad s_{3}=\left(p_{1}+p_{2}\right)^{2},\\
s_{1}+s_{2}+s_{3} & = & M_{\eta'}^{2}+M_{\eta}^{2}+2m_{\pi}^{2}.
\end{eqnarray*}
Roy equations and chiral perturbation theory allow one to describe accurately the amplitude of $\pi\pi$ scattering in low energy region, that is allowed in considered here decay. This approach makes it possible to control the analytical continuation of the amplitude to the complex plane, where the pole interpreted as $\sigma$-meson is observed  \cite{roy}.

Due to unitarity the amplitude of $\pi\pi$-scattering should obey rather strong conditions. First of all, for $s<\left(2m_{\pi}\right)^{2}$ this amplitude should be real. For $s\ge\left(2m_{\pi}\right)^{2}$ up to $KK$-production threshold the imaginary part of this amplitude should be
\begin{eqnarray*}
\mbox{Im}\left(\frac{1}{\mathcal{A}_{\pi\pi}(s)}\right) & \sim & \sqrt{1-\frac{4m_{\pi}^{2}}{s}}.
\end{eqnarray*}
Chiral perturbation theory, in addition, that at $s=s_{A}=m_{\pi}^{2}/2$ the amplitude equals to zero (Adler "self-consistency" conditions \cite{Adler}).

The amplitude that satisfied listed above conditions can be expressed through the series over the variable
\begin{eqnarray*}
w\left(s\right) & = & \frac{\sqrt{s}-\sqrt{4m_{K}^{2}-s}}{\sqrt{s}+\sqrt{4m_{K}^{2}-s}},
\end{eqnarray*}
that transforms couples $s$-plane with cuts $s\le0$ and $s\ge\left(2M_{K}\right)^{2}$ into a disc $\left|w\right|<1$ in complex $w$-plane (it is clear that $w\left(4M_{K}^{2}\right)=1$, $w(0)=-1$). The introduction of a new variable improves the convergence of the series in the considered variable domain. The amplitude of $\pi\pi$-scattering can be written in the form
\begin{eqnarray}
\mathcal{A}_{\pi\pi}(s) & \sim & \kappa t_{0}^{0}(s)=\kappa\left\{ \frac{m_{\pi}^{2}}{s-s_{A}}\left[\frac{2s_{A}}{m_{\pi}\sqrt{s}}+B_{0}+B_{1}w\left(s\right)+\cdots\right]-i\sqrt{1-\frac{4m_{\pi}^{2}}{s}}\right\} ^{-1},\label{eq:ampPiPi}
\end{eqnarray}
where $\kappa$ is unknown constant that can be determined from the fit of experimental data.

The analysis of NA48/2 results in $K_{e4}$ decay \cite{na48} tells us, that in the presented above series one can leave only first two terms with the coefficients
\begin{eqnarray*}
B_{0} & = & 7.4,\qquad B_{1}=-15.1
\end{eqnarray*}
These values correspond to following position of $\sigma$-meson pole:
\begin{eqnarray*}
\sqrt{s} & = & (459 + 259 i)\,\mbox{MeV}
\end{eqnarray*}

Let us now proceed to $\pi\eta$ channel. In our kinematic region the main contribution in this channel comes from virtual $a_{0}$-meson. It is well known that its mass is close to $KK$ production threshold. As a result the propagator of this meson is different from usual Breit-Wigner form: one should introduce the self-energy corrections caused by $\pi\eta$, $\pi\eta'$ and $KK$ loops. It should be noted, that these corrections are significant at the pole. Our analysis, however, show, that in our kinematical region these corrections are small and are above current experiments accuracy. The form of $a_{0}\pi\eta$ and $a_{0}\pi\eta'$ vertices, on the other hand, is important.

These vertices can be written in several forms. First of all one can use simple point-like representation with effective constants $g_{\pi\eta}$ and $g_{\pi\eta'}$. The first constant can be determined from experimental value of $a_{0}$-meson width:
\begin{eqnarray*}
\Gamma\left(a_{0}\to\pi^{0}\eta\right) & = & \frac{g_{\pi\eta}^{2}}{8\pi m_{a}^{2}}\left|\mathbf{p}\right|\approx\Gamma_{a_{0}}=50\div100\,\mbox{MeV},
\end{eqnarray*}
so
\begin{eqnarray*}
1.95\,\mbox{GeV} & < & g_{\pi\eta}<2.75\,\mbox{GeV}.
\end{eqnarray*}
This value agrees with $g_{\pi\eta}=(2.46\pm0.08\pm0.11)$ GeV, presented in recent experimental work \cite{Ambrosino:2009py}
The constant $g_{\pi\eta'}$ can be determined either from $SU(3)$-symmetry relations \cite{Feldmann:1998sh} or directly from fit of the considered in this article process. It should be noted, that $SU(3)$ relations require information on quark structure of $a_{0}$-meson, that is widely discussed question. For this reason we use experiment for determining this constant.

Another form of interaction vertex is motivated by chiral perturbation theory. According to it the vertex should be equal to zero in the limit $p_{\pi}\to0$. In this case the vertex can be written in the form $\left(p_{\pi}p_{\eta}\right)\gamma_{\pi\eta}$ for $a_{0}\to\pi\eta$ interaction and $\left(p_{\pi}p_{\eta'}\right)\gamma_{\pi\eta'}$ for $\eta'\to a_{0}\pi$ interaction. This form was used in paper \cite{schechter}. It seems more attractive, since in $\pi\pi$ scattering amplitude chirality conditions are taken into account. The constant $\gamma_{\pi\eta}$ can be determined from $a_{0}\to\pi\eta$ decay width
\begin{eqnarray*}
5.7\,\mbox{GeV}^{-1} & < & \gamma_{\pi\eta}<8.1\,\mbox{GeV}^{-1},
\end{eqnarray*}
while for determination of $\gamma_{\pi\eta'}$ constant one can use $SU(3)$-symmetry or distribution of $\eta'\to\eta\pi\pi$ decay over Dalitz region.

Thus, we will use following two variants of $\pi\eta$-scattering amplitude:
\begin{eqnarray}
\mathcal{A}_{\pi\eta}(s) & = & \frac{g_{\pi\eta}g_{\pi\eta'}}{s-m_{a}^{2}+i\Gamma(s)m_{a}}\label{eq:APiEtaSimple}
\end{eqnarray}
or
\begin{eqnarray}
\mathcal{A}_{\pi\eta}(s) & = & \gamma_{\pi\eta'}\gamma_{\pi\eta}\frac{\left(p_{\eta}p_{\pi}\right)\left(p_{\eta'}p_{\pi}\right)}{s-m_{a}^{2}+i\Gamma(s)m_{a}}.\label{eq:ApiEtaChir}
\end{eqnarray}

\section{Fit Results}

Our parametrization has following unknown constants: coupling constant of $\sigma$-meson with $\eta'\eta$ pair interaction $\kappa$ (see formula \eqref{eq:ampPiPi}) and constants of $a_{0}$-meson interaction with $\eta\pi$ and $\eta'\pi$ states ($g_{\pi\eta}$, $g_{\pi\eta'}$ or $\gamma_{\pi\eta}$, $\gamma_{\pi\eta'}$ depending on the form of $a_{0}$ vertices). These constants will be determined using the value of $\eta'\to\eta\pi^{0}\pi^{0}$ decay width and fit of the Dalitz plot of this decay, obtained in GAMS-$4\pi$ experiment \cite{neutral}.

\begin{figure}[htb]
\begin{center}
\begin{picture}(440,440)
\put(0,0){\includegraphics{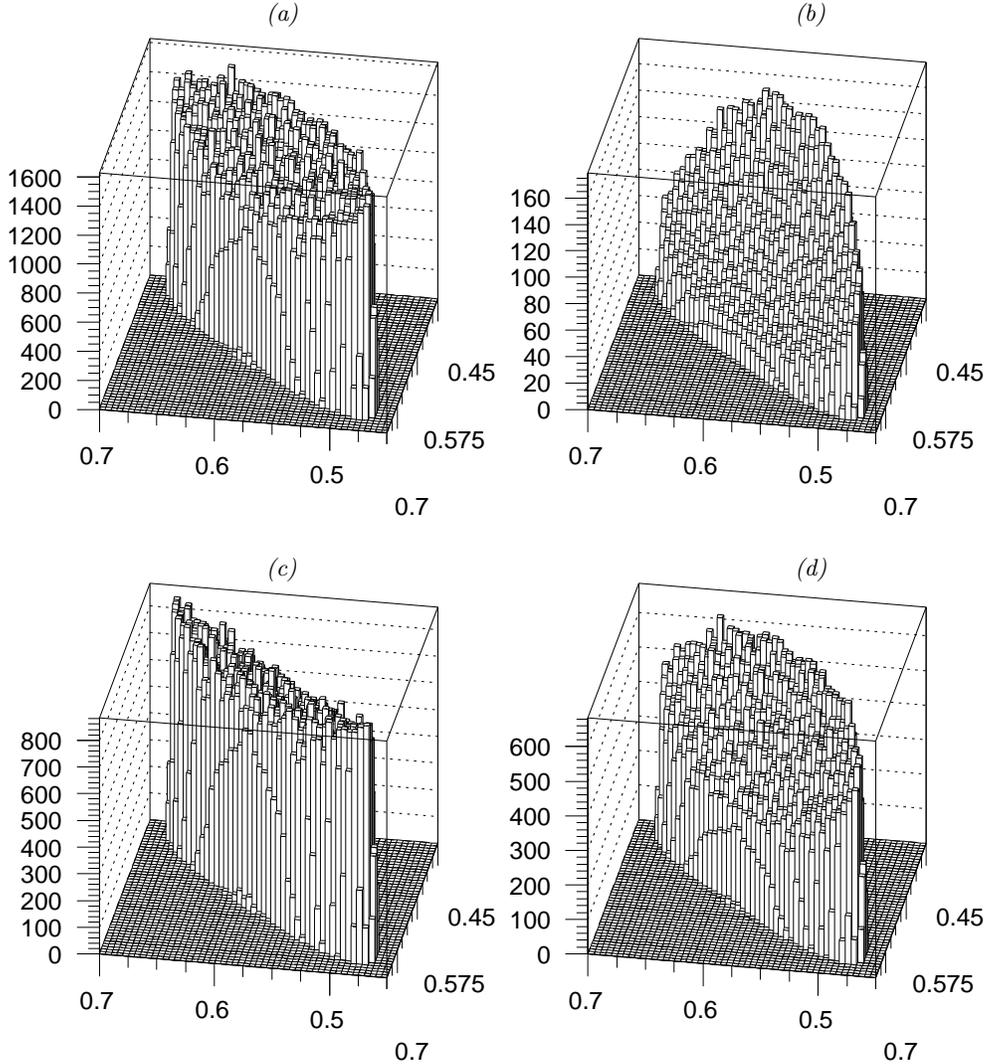}}
\put(120,405){\it (a)}\put(320,405){\it (b)}
\put(120,195){\it (c)}\put(320,195){\it (d)}
\end{picture}
\end{center}
\caption{%
(a) --- experimental Dalitz distribution for the decay (\ref{epp}) from  \cite{neutral};
(b) --- MC results with only $\sigma$-meson taken into account;
(c) --- MC results with only $a_0$-meson taken into account;
(d) --- $a_0-\sigma$-interference
\label{fig:Dalitz}
}
\end{figure}

Let us first consider the point-like interaction of $a_{0}$-meson \eqref{eq:APiEtaSimple}. Best agreement with experimental data (with confidence level CL=0.52) is observed when coupling constants are equal to
\begin{eqnarray*}
\kappa & = & -4.0\pm0.3,\\
g_{\pi\eta}g_{\pi\eta'} & = & (0.93\pm 0.3)\ \mathrm{GeV}^2.
\end{eqnarray*}
Above it was mentioned that coupling constant $g_{\pi\eta}$ is connected with $a_{0}$-meson decay width ($2\,\mbox{GeV}<g_{\pi\eta}<3\,\mbox{GeV}$). As a result we get following restrictions on $g_{\pi\eta'}$ coupling constant:
\begin{eqnarray*}
0.36\,\mbox{GeV} & < & g_{\pi\eta'}<0.51\,\mbox{GeV}.
\end{eqnarray*}
If we assume that $a_{0}$-meson is build from light quarks only ($a_{0}\sim u\bar{u}+d\bar{d}$), these constants should be connected from $\eta-\eta'$ mixing. For example, in quark mixing scheme \cite{Feldmann:1998sh} with mixing angle $\Phi\approx40^{0}$ the ratio of these constants should be \begin{eqnarray*}
\frac{g_{\pi\eta'}}{g_{\pi\eta}} & = & \tan\Phi=0.8
\end{eqnarray*}
It can be easily seen, that this relation does not hold for presented above values of coupling constants $g_{\pi\eta}$ and $g_{\pi\eta'}$. This fact is not surprising, on the other hand, since current information on internal structure of $a_{0}$-meson is rather poor. It is, for example possible, that there is noticeable $s\bar s$ component in this meson, as it was mentioned in \cite{Ambrosino:2009py}.

The situation is different when one use chiral form of $a_{0}$ vertices (expression \eqref{eq:ApiEtaChir}). In this case best agreement with experiment (CL=0.92) is observed at
\begin{eqnarray*}
\kappa & = & -4.0\pm0.3\\
\gamma_{\pi\eta}\gamma_{\pi\eta'} & = & (35\pm 4)\,\mbox{GeV}^{-2}.
\end{eqnarray*}
Form experimental width of $a_{0}$-meson one can determine the value of $\gamma_{\pi\eta}$-constant ($5.7\,\mbox{GeV}^{-1}<\gamma_{\pi\eta}<8.1\,\mbox{GeV}^{-1}$). It is easy to obtain the following restrictions on coupling constant $\gamma_{\pi\eta'}$:
\begin{eqnarray*}
4.6\,\mbox{GeV}^{-1} & < & \gamma_{\pi\eta'}<6.6\,\mbox{GeV}^{-1}.
\end{eqnarray*}
The relation caused by $SU(3)$-symmetry holds for these constants significantly better, then in the case of point-like $a_0\pi\eta$ interaction. We would like to note, that these values are close enough (up to a sign) to results of the work \cite{schechter} where $\gamma_{\pi\eta} = 6.8 \div 7.2\ \mathrm{GeV}^{-1}$, $\gamma_{\pi\eta'} = 7.4 \div 8\ \mathrm{GeV}^{-1}$. It is clear, that difference in sign is insignificant, since one freely change the total sign of the amplitude.

\section{Discussion}

In fig.\ref{fig:Dalitz} we present experimental Dalitz plot of $\eta'\to\eta\pi^{0}\pi^{0}$ decay (fig.\ref{fig:Dalitz}a), and the results of Monte-Carlo modeling with only $a_{0}$-resonance contribution taken into account (fig.\ref{fig:Dalitz}b), only $\sigma$-meson contribution taken into account (fig.\ref{fig:Dalitz}c), and only interference between $a_{0}$ and $\sigma$ (fig. \ref{fig:Dalitz}d). From these figures it is clearly seen, that the $a_{0}$-meson gives the main contribution. The contribution of $\sigma$-meson is significantly smaller, while the role of $a_{0}-\sigma$ interference is comparable with that of $a_{0}$-meson.

It is interesting to note, that, in spite of dominant role of $a_{0}$ resonance, $Y$-distribution generated by $a_{0}$-resonance only have opposite slope in comparison with experimental data. After inclusion of $\sigma$-meson the agreement with the experiment is restored. So we can conclude, that, though the contribution of $\sigma$-meson to partial width of the decay $\eta'\to\eta\pi\pi$ is small, it plays a crucial role in description of experimental data.

\begin{figure}
\begin{centering}
\includegraphics{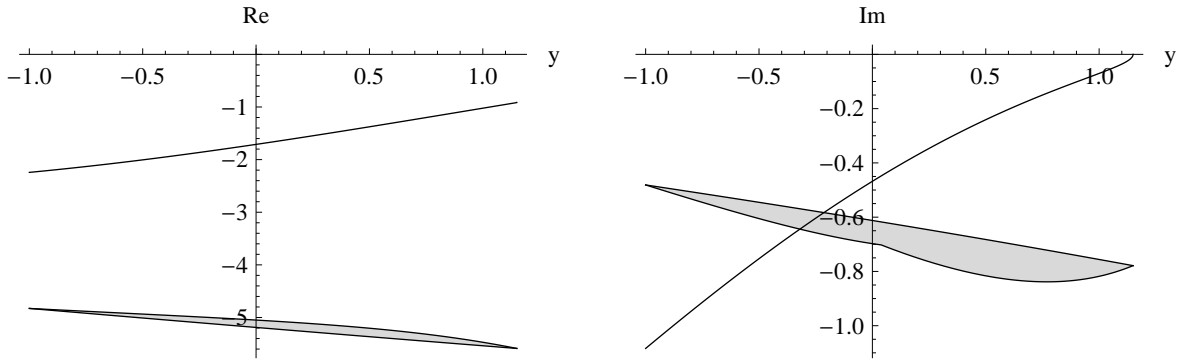}
\end{centering}
\caption{Real and imaginary parts of the amplitudes of $\sigma$-meson (thin line) and $a_0$-meson (dashed region)\label{fig:ReIm}}
\end{figure}

This point can be clearly seen from studying contributions of different resonances to real and imaginary parts of the amplitude. In fig \ref{fig:ReIm} we show real and imaginary parts of the amplitudes of $a_{0}$ and $\sigma$ exchange versus Dalitz variable $Y$. In our parametrization $\sigma$-meson amplitude is $X$ independent, what can be seen from thin lines on these distributions. Amplitude of $a_{0}$-exchange, on the other hand, depends on $X$, so one can see a number of lines (each individual line corresponds to different $X$ values). Amplitudes shown in fig.\ref{fig:ReIm} agrees well with fig.11, presented in paper \cite{schechter}. Difference is caused by difference in parameterizations of resonance amplitudes.  although we use slightly different parameterizations for $\pi\pi$- and $\pi\eta$-rescattering amplitudes. From this figure it is clear, that the contribution of $a_{0}$-meson exchange dominates in the real part of the amplitude, while in the imaginary part main contribution comes from $\sigma$-meson exchange.

\begin{figure}
\begin{centering}
\includegraphics[scale=0.8]{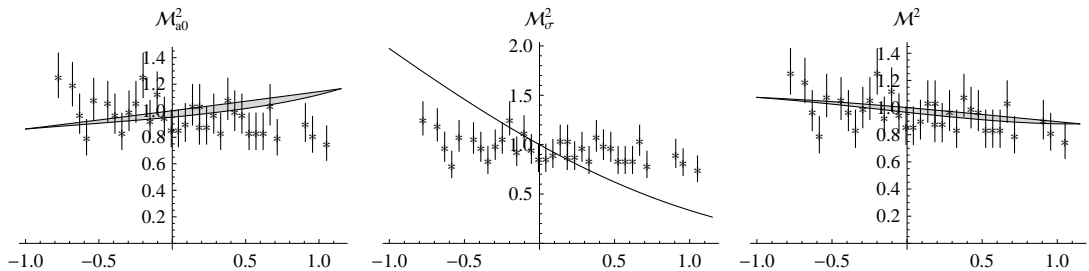}
\par\end{centering}
\caption{Squared matrix element as a function of Dalitz variable $Y$. Plots (from left to right) correspond to $a_0$-meson only, $\sigma$-meson only, and total matrix elemet\label{fig:Matr2}}
\end{figure}

In order to demonstrate crucial role of $\sigma$-meson for description of Dalitz distribution it is useful to trace contributions of different mesons with presented above parameters into squared matrix element. In figure \ref{fig:Matr2}a we show $Y$-distribution with only $a_{0}$-meson exchange taken into account (each line in this graph corresponds to specific $X$ value), in fig.\ref{fig:Matr2}b $Y$-distribution with only $\sigma$-meson taken into account (this amplitude does not depend on $X$, so only one thin line is present), while in fig.\ref{fig:Matr2}c all terms of the amplitude are used. It can be clearly seen from this figure, that experimental slope of the $Y$-distribution can be achieved only if $\sigma$-meson is included. We think that this proves firmly the necessity of $\sigma$-meson inclusion for analysis of the considered process $\eta'\to\eta\pi^0\pi^0$.

\section{Conclusion}

In our work we analyze experimental data of the reaction $\eta'\to\eta\pi^0\pi^0$ in the framework of isobar model taking into account contributions of scalar mesons. We show, that it is sufficient to include $a_{0}$- and $\sigma$-mesons to describe experimental data accurately.

In the $a_{0}$-meson case one can use simple Breit-Wigner parametrization with energy-dependent width. We consider two different forms of $a_{0}\to\eta\pi$ and $a_{0}\to\eta'\pi$ vertices and compare corresponding coupling constants with relations caused by $SU(3)$-symmetry. Our analysis show that in the case of chiral-type interaction coupling constants agrees $SU(3)$-symmetry much better that point-like couplings.

The situation is more complicated for $\sigma$-meson, since the width of this meson is comparable with its mass, and simple Breit-Wigner parametrization cannot be used. For this reason a more accurate description of $\sigma$-meson exchange amplitude is required. This amplitude should satisfy a number of conditions: analiticity, unitarity, and crossing symmetry. In our article we use $\pi\pi$-amplitude, obtained from fit of the experimental data of $K_{e4}$-decay \cite{na48}. It contains a pole in complex plane at $\sqrt{s}=(459+259 i)$ MeV, that is associated with $\sigma$-meson.

We show, that, in agreement with previous work \cite{schechter}, $\pi^{0}\eta$-rescattering via virtual $a_{0}$-meson gives main contribution to partial width of the $\eta'\to\eta\pi^{0}\pi^{0}$ decay. For description of the Dalitz distribution, on the other hand, it is necessary to take into account $\sigma$-meson contribution and its interference with $a_{0}$ one. If these effects are neglected, the slope of the Dalitz distribution in the variable $Y$ is opposite to experimental value.$ $Inclusion of the $\sigma$-meson corrects the situation. It should be noted, that $\sigma$-meson is very exotic particle, that has width comparable with its mass. For this reason direct experimental observation of this particle is very problematic. We think that presented in this article analysis of $\eta'\to\eta\pi^{0}\pi^{0}$ decay gives additional argument in favor of $\sigma$-meson existence.

\section*{Acknowledgments}

The authors would like to thank A.M. Zaitsev for pointing our attention to work \cite{Deshpande:1978iv}. This work was financially supported by Russian Foundation for Basic Research (grants \#09-02-00132-a and 07-02-00417-a). One of the authors (A.V.L.) was also supported by President grant (\#MK-110.2008.2), grant of Russian Science Support Foundation and noncommercial foundation "Dynasty".

\end{document}